# *A New Trusted and Collaborative Agent Based Approach for Ensuring Cloud Security*

[1]Shantanu Pal, [2]Sunirmal Khatua, [3Corresponding Author]Nabendu Chaki, [4]Sugata Sanyal

[1, 2, 3]92 A. P. C. Road, University of Calcutta, India
[4]Tata Institute of Fundamental Research, India.

[1]*shantanu.smit@gmail.com;* [2]*enggnimu_ju@yahoo.com;* [3]*nabendu@ieee.org;* [4]*sanyal@tifr.res.in*

**Abstract:** In order to determine the user's trust is a growing concern for ensuring privacy and security in a cloud computing environment. In cloud, user's data is stored in one or more remote server(s) which poses more security challenges for the system. One of the most important concerns is to protect user's sensitive information from other users and hackers that may cause data leakage in cloud storage. Having this security challenge in mind, this paper focuses on the development of a more secure cloud environment, to determine the trust of the service requesting authorities by using a novel VM (Virtual Machine) monitoring system. Moreover, this research aims towards proposing a new trusted and collaborative agent-based two-tier framework, titled *WAY* (Who Are You?), to protect cloud resources. The framework can be used to provide security in network, infrastructure, as well as data storage in a heterogeneous cloud platform. If the trust updating policy is based on network activities, then the framework can provide network security. Similarly, it provides storage security by monitoring unauthorized access activities by the Cloud Service Users (CSU). Infrastructure security can be provided by monitoring the use of privileged instructions within the isolated VMs. The uniqueness of the proposed security solution lies in the fact that it ensures security and privacy both at the service provider level as well as at the user level in a cloud environment.

## 1. Introduction

In recent times, cloud computing is evolving as a revolutionary technique for the way we compute. In its way of evolution, starting from cluster computing through grid computing, it considered two important parameters of distributed computing paradigm namely flexibility and utilization. While cluster computing provides high flexibility of managing the resources at the cost of lower resource utilization and grid computing provides better utilization of resources at the cost of lesser flexibility of managing those resources, cloud computing provides both high flexibility as well as high resources utilization. However we are gaining those advantages at the cost of high security threats and privacy challenges since cloud computing deals with the computation and data at third party's infrastructure.

Cloud computing deals with providing storage and computation resources as a service to the Cloud Service Users (CSU) in the form of Software as a Service (SaaS), Platform as a Service (PaaS), Infrastructure as a Service (IaaS), Storage as a Service (SaaS) etc. Software as a Service [3] ensures to provide services as pay-as-you-go pricing scheme where customer does not need to install configure or run the application on their local computers. Platform as a Service offers a software execution environment to deploy Web-based applications. Users do not need to think about the cost and complexity of buying servers or setting the infrastructure. Therefore PaaS refers to provide a development platform to deploy, host or maintains their applications. Infrastructure as a Service shares hardware resources for executing services using virtualization.

To date from a small investor to a big IT company everyone is now relying on this system. Cloud computing has several advantages such as ease to use and maintenance, need low power consumption for operation and reductions in the overhead for storing and servicing the data. In spite of several advantages cloud also suffers from different security threats and risks to protect its resources from unauthorized users and hackers. These security threats and attacks are the biggest concern towards the improvement of a more secure cloud infrastructure. Traditional mythologies are not enough to adopt for protecting cloud resources as they become obsolete with respect to the ever evolving security threats as well as to avoid data losses in the cloud environment. Moreover data stored in cloud is not just merely stored, but rather this data gets accessed by large number of times and changes in the form of insertion, deletion or updation that take place from time to time.

Security and privacy even with traditional information security systems and networks has been difficult to satisfy and this is also a challenging job for cloud environment [1]. The primary focus of this paper is to introduce a novel and trusted security framework for securing cloud resources. The rest of the paper is organized as follows: Section 2 provides a state of the art security review in cloud computing environment. Section 3 describes the proposed security model in details. Section 4 analyzed the performance of the proposed scheme in a simulated test bed. Finally, in section 5 the concluding comments are drawn.

## 2. State of the Art Review on Security for Cloud Environment

Issue of data security is mentioned as the biggest factor for the cloud computing [2]. Security challenges in a cloud computing environment may be classified as: (1) Protection of data towards user's side. (2) Protection of data towards service provider's end and (3) Protection of data in storage server or Cloud Data Center (CDC).

Most of the security models so far are based on traditional cryptographic approaches. Some referred to dynamic security measures for a cloud environment [3] while other domain based applications discussed the growing security concerns of cloud infrastructure. In [2] domain trust concept is used to develop a secure cloud infrastructure. In [4], Jin-Song Xu et al, separates contain and format from documents, before handling and storing of data in to the remote cloud data center to protect cloud resources from unauthorized users or hackers. An optimized authorization method (using encryption functions) is used for accessing data base for trusted CSUs.

In Software as a Service, SQL injection attacks are one of the most important factors to be considered. In a SQL injection attack [5], an attacker introduces malicious codes with the standard SQL codes. In this way attackers get informed about the sensitive information from secure databases. This is an important concern for the cloud storage from non-trusted CSUs. Another important attack known as Cross Site Scripting (XSS), in which an attacker injects malicious scripts into the Web contents. If any user clicks on them, the sensitive information will automatically redirects to the attacker machine. In cloud computing this may happen if any attacker affects the cloud service interface, as cloud application should requires Web interfaces for their services.

Domain Name Service (DNS) attacks routed the secure web pages to the non-trusted users [6]

where any secure domain redirects to any of those CSUs. Attackers are also hacks sensitive information from any secure domain using the time lag for a reuse of any IP (Internet Protocol) address. In cloud domain address are changing every time for any specific CSU, hence this attacks does malicious activities in cloud storage.

In Denial of Service Attacks (DoS) attackers sends a huge number of requests are sent by the attackers, thus the actual service become unavailable to the trusted users. In Distributed Denial of Service (DDoS) attack, in order to deny the important services running on a server, the destination sever is flooded with an umpteen number of packets such that the target server is not able to handle it.

## 3. Proposed Cloud Security Framework

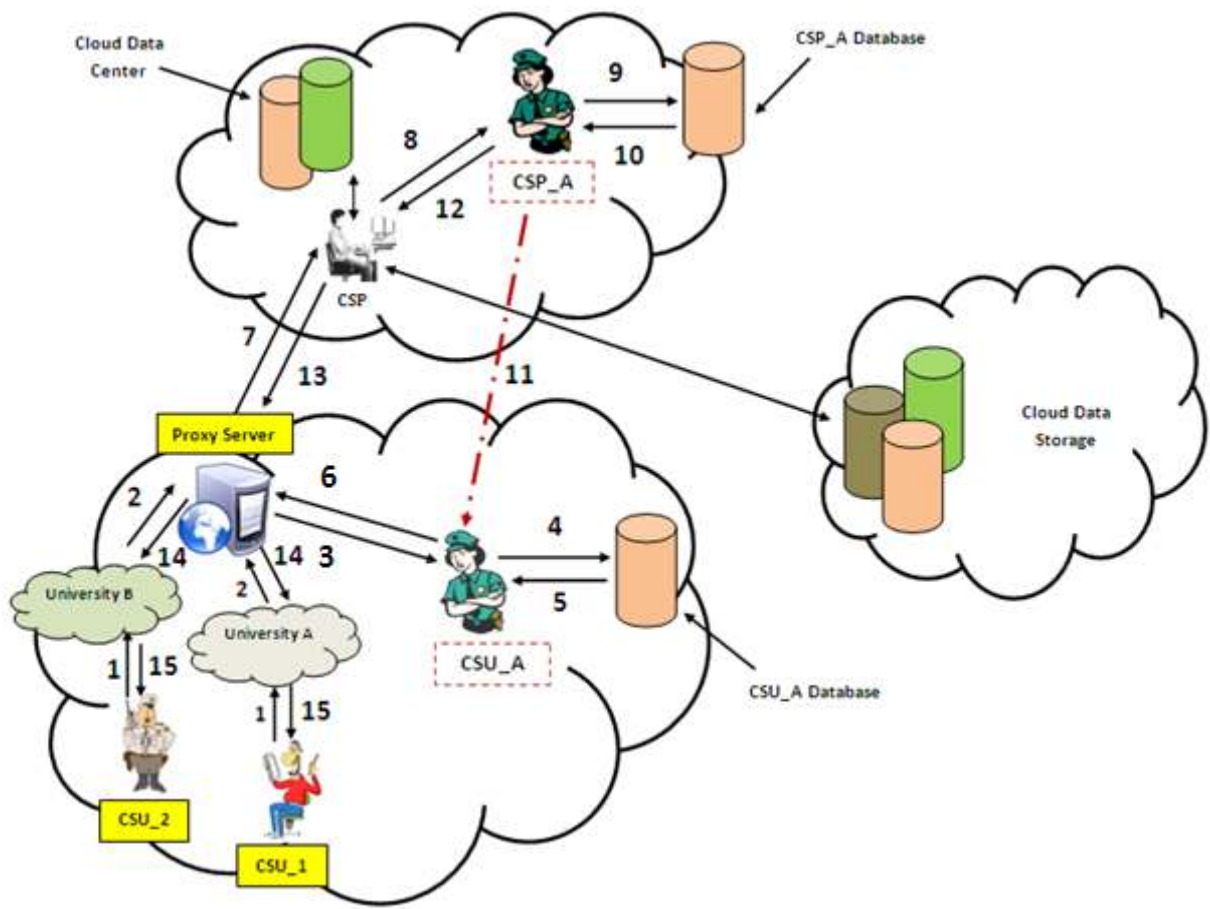

Figure 1. Proposed Cloud Security Framework

The term '*WAY*' denotes a way of secure data communication between the Cloud Service Provider (CSP) and the Cloud Service User (CSU) in a heterogeneous cloud computing environment. It is done by asking *'Who Are You?'* to determine and satisfy the basic trust requirements of the service requesting CSUs. Two-tier architecture has been proposed in this paper. One is *Broker Domain* and another is *Cloud Service Provider Domain* where Broker Domain is denoted by Level_1 and Cloud Service Provide Domain is denoted by Level_2. This two level communication authenticates the trusted CSUs for accessing private information from cloud data storage.

The system architecture is presented in Figure 1. Proposed novel VM monitoring techniques assures the trustworthiness of the system by calculating current or updated trust degree for each service requesting CSU and the domain from where the request is coming. In this model CSU (such as CSU_1 in Figure 1) requests information from the Cloud Service Provider (CSP). As the first step, when any CSU who wants to send any request to the CSP, they have to pass the correct authentication data (such as user id and password set for them by the system) through a proxy server situated in its domain. In this approach proxy server is used as a communication channel between two domains. As an example, University A and University B is denoted for a specific group of users who requires their University specific authentication data for sending their requests to CSP through the proxy server. When the request passes through the proxy server it reaches to the trusted-agent situated at broker domain (denoted as Cloud Service User Agent, simply CSU_A in Figure 1). The trust degree of the service requesting CSU is then checked by this CSU_A with its local databases presents in the same domain (in Broker Domain or in Level_1). If the current (or updated) trust value for this CSU is greater than the threshold value set for this communication, then only the request reaches to the CSP (situated in CSP Domain or in Level_2).

When this request for information reaches to the CSP, it's immediately passes the request to the trusted-agent (denoted as Cloud Service Provider Agent, simply CSP_A) situated in the same domain to check the trust degree of the domain from where this request came. Agent CSP_A then checks the current or updated (this updation is done by the previous successful of unsuccessful iteration information) trust degree for this particular domain and then sends this result back to the CSP, only if trust degree is greater than the current threshold value set for this domain. Then the CSP will allow passing the requested information back to that particular CSU through the proxy server. CSP_A will update the trust degree after successfully executing the task. If the domain trust is less than that of the current or updated trust degree, or the CSU does any malicious activities, CSP_A immediately inform this report to the CSU_A situated at broker domain for taking necessary actions. CSU_A will in turn decrease the trust value for this particular user. After few (depending on the types of communications) non-trusted activities or reports, CSU_A will remove this particular CSU from its domain.

One of the major advantages of this framework is that, domain remains unaffected (with only decreased amount of trust degree than that of non-trusted users) when a said non-trusted CSU does malicious activities in the system. The trust degree of the domain will decrease accordingly with the malicious activities and updating policies. The CSP_A and CSU_A maintain their own databases, user activities information and updated trust degrees for calculating updated trust degree.

The flow of information as denoted in Figure 1 is discussed below to understand the process of the proposed model.

**Step 1:** CSU provides credentials to the proxy server to request an information (using credentials set for this University).

**Step 2:** Authentication information of this user reaches to the proxy server. If this information is correct, request will reaches to the CSU_A similarly for any incorrect information request will be dropped.

**Step 3:** Request came to CSU_A for checking current (or updated) trust degree for this CSU.

**Step 4:** The CSU_A checks the trust degree with its database.

**Step 5:** Update result.

**Step 6:** If the trust degree of the service requesting CSU is greater than the threshold value, CSU_A sends this request to CSP via proxy server.

**Step 7:** Request reaches to the CSP.

**Step 8:** CSP passes request to the CSP_A for checking the trust degree of the domain from where the request came.

**Step 9:** CSP_A checks it local databases.

**Step 10:** Update result.

**Step 11:** If the domain's trust degree is not greater than the threshold trust degree set for this domain, the request will be dropped and update this information to CSU_A for further action.

**Step 12:** Request for information satisfies the basic trust degree.

**Step 13:** CSP will send data to the service requesting CSU via proxy server.

**Step 14:** 'Data' determines and reaches to the correct CSU through the proxy server.

**Step 15:** CSU get the desired data.

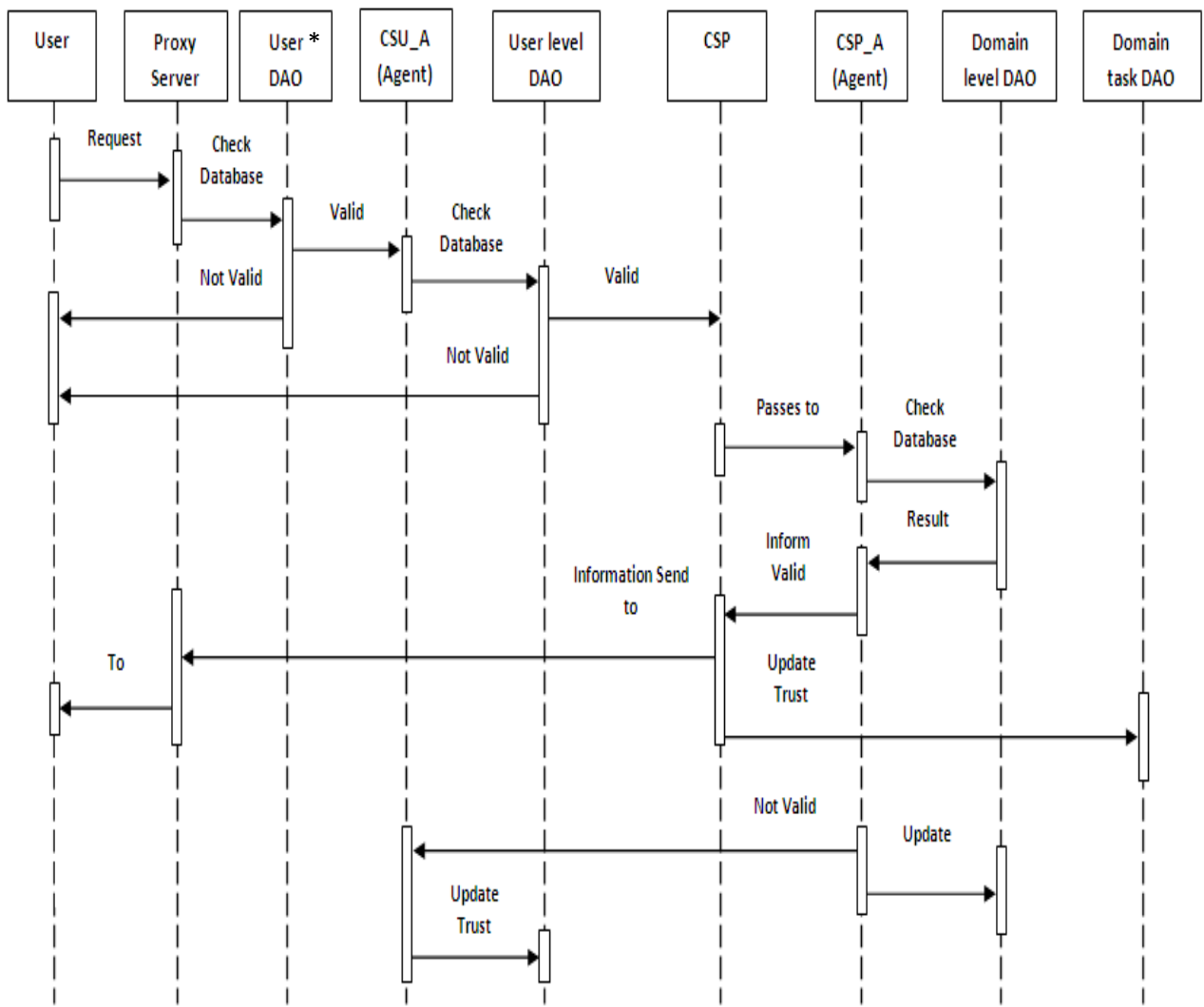

*DAO represents database

**Figure 2. Sequence Diagram of the Proposed Framework**

## 3.1 Architecture and Design

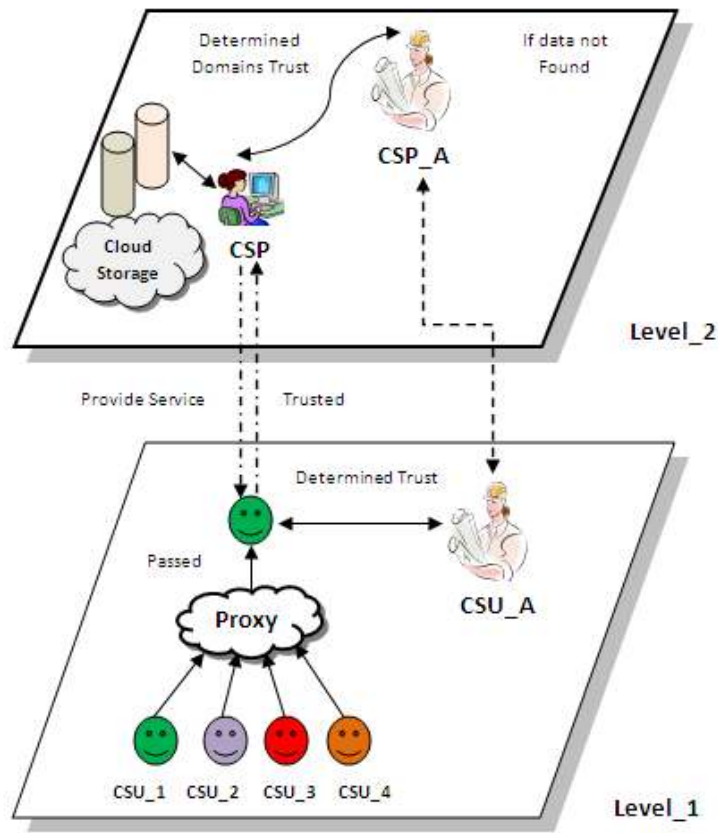

**Figure 3. Proposed Two-Tier Security Architecture for Cloud Computing**

The proposed two-tier security framework for cloud along with its components has been introduced in this section. The proxy server in tier 1 and CSP in tier 2 together maintain the cloud security.

**Cloud Service User (CSU):** Request services from cloud storage.

**Cloud Service Provider (CSP):** Provides services to the requesting trusted CSU.

**Proxy Server:** To check the preliminary authentication information set for the CSU of a particular domain and maintain the way of communication.

**Cloud Service User Agent (CSU_A):** Checks the trust degree set for a service requesting CSU and updates the corresponding trust degree, after performing one or more successful or unsuccessful communication.

**Cloud Service Provider Agent (CSP_A):** Checks the domain's trust degree of the service requesting CSU.

## 3.2 Algorithms

In this section, two proposed algorithms are discussed to determine the trustworthiness of any service requesting CSU and getting information if the request satisfies the basic trust requirements.

The algorithm for requesting information operates for a cloud service user $\eta$ that sends request $\Gamma$ to cloud service provider Ś for an information $\xi$. The user id $I$, password $p$ and the current trust value $v$ are to be passed as input arguments to the algorithm. The algorithm works for all three types of users; i.e., user could be trusted, innocent or non-trusted. It may be noted that this algorithm would be hosted in the proxy server, say P.

**Algorithm for Requesting Information**

***Input:*** *User id I; User password p; User Trust Value v; service request $\Gamma(\xi)$*

***Output:*** *$\Gamma$ to be accepted or rejected*

Begin Algorithm

Proxy server *P* Checks *(I, p);*
*If validation succeeds*
   *P passes the request to Cloud Service User Agent (CSU_A);*
   *CSU_A Checks v;*
   *If v > threshold*
      *Request $\Gamma$ for information $\xi$ reaches Ś;*
*else*
   *Drop the request $\Gamma$;*
End.

The second algorithm is for delivering information. This would be operational when a successful request $\Gamma$ reaches to Cloud Service Provider Ś for an Information $\xi$ from domain D. The trust value for the user would be revised accordingly.

**Algorithm for Delivering Information**

***Input:*** *Cloud service request $\Gamma$, trust value V for domain D*

***Output:*** *Delivering information $\xi$ to user i or reject service*

Begin Algorithm

Ś sends incoming request for $\Gamma$ to Cloud Service Provider Agent (CSP_A);
If V> threshold
Send $\xi$ to user I via Proxy Server P;
CSP updates the database for the trust value;
else
   Drop the request $\Gamma$;
   CSP_A inform CSU_A about I;
   *CSU_A updates databases for I;*
End.

## 3.3 Simulation and Experimental Results

In a cloud computing environment different services are carried out on behalf of customers on hardware to which the customers have no access. The input data for cloud services is uploaded by the user to the cloud storage that means they typically result in user's data being present in unencrypted form on a machine that the user does not own or control [7]. This poses some inherent challenges in terms of security and privacy for the system where one of the top risks is the delivery of private data to an unauthorized user. The basic motivation of developing this proposed architecture is to stop the services of a non-trusted CSU in a heterogeneous cloud environment after unsuccessfully executing any request.

For simulation three types of CSU have been considered, they are as Trusted, Innocent and Non-Trusted. There are two types of tasks they may be carried out automatically during the communication and they are denoted as trusted task and non-trusted task. There are several requests that can be processed in a certain interval of time to perform these tasks. User id, Task id and Domain id for corresponding CSU, task and domains are given in the sytem. Simulation code is written in Java programming language, the performance analysis is done on a computer having following configuration: 2 GB RAM, 500 GB Hard disk an Intel core i3 processor @2.27 GHz.

### 3.4 Assumptions for this simulation

CSU's trust is the user's identity trust. User's identity trust in cloud is not enough, the issues of user's behavior trust should also be evaluated and managed. So it needs a mutual mechanism to establish trust between the CSUs and the CSPs as (1) user's trust to the provider and (2) provider's trust to the users.

A trusted monitoring function should be integrated into the system to supervise the participant CSU's behavior and depending on user's behavior a trust management mechanism must be incorporated to update CSU's trust value.

After evaluating user's behavior, it is required to manage thistrust value efficiently. Proposed novel trusted and collaborative agent-based security framework takes users behavior evidence from CSP and manage user trust value from it.

### 4. Performance Analysis

The proposed security framework is based on a trust model. The trust degree of any CSU increases after performing any trusted communication. Similarly for any non-trusted communication the trust degree decreases and the corresponding trust table updated by the trusted-agents for next task. The probability of executing a task successfully for any non-trusted user is very low than that of any trusted and innocent users in the same domain.

A novel trust-based algorithm [8] is used to determine the trust of any service requesting user to deliver the requested information from cloud data storage. Suppose there are U numbers of users present in a domain D where U = $\{u_1, u_2, u_3,\ldots, u_n\}$. These users may be trusted, innocent or non-trusted. U Є $\{T_1, T_2, T_3\}$. Where $T_1$ represent trusted user, $T_2$ represent innocent user and $T_3$ represent non-trusted user. Similarly there are two kind of tasks that may be performed during any communication; they are trusted task, denoted by $T_t$ and non-trusted task denoted by $T_n$. There are *N* numbers of tasks (where N Є $\{n_1,n_2,n3,\ldots,n_n\}$) that can be performed in a simulation. For any instance a user $u_1$ belongs to $T_1$ can perform the task of type $T_t$ for $n_1$ times.

As for example, a trusted user after successfully completion of a given task T the trust degree will be increased and after an unsuccessful communication the trust degree will be decreased accordingly to the performance. Probability function is used to determine the trust degree of any service requesting CSU and then marked them as trusted, innocent or non-trusted. It should be noted that the probability of getting higher trust value by performing a trusted task by a trusted user, is always better than a non-trusted or an innocent user.

Actions for any task can be positive or negative. However, it is not assumed that all negative actions are not the same that is the reason because we distinguish between wrong actions and malicious actions [8]: *Positive*, i.e. right actions done by the trusted user; *Wrong*, i.e. bad actions that do not cause any damage or may cause damages done by the innocent user; and *Malicious*, i.e. harmful actions such as attacks done by the non-trusted user.

Accesses to authorized resources and suitable use of them are considered right actions. An entity can make wrong actions by mistake or intentionally, but it is difficult to know.

To calculate the *action value* $V_a$, we take into account the performed action weight, but this value is penalized or rewarded by the past behavior. This function increases or decreases according to the performed positive and negative actions respectively. The equation is denoted as follows:

$$V_a = \left(1 - \frac{A_N}{Total_a}\right) \cdot W_a^{(m)}$$ ............... (1) Where $0 \leq V_a \leq 1$

In equation (1), $\left(1 - \frac{A_N}{Total_a}\right)$ represents the past behavior of any CSU. This value tends to 0 when the behavior is negative, and it tends to 1 when the behavior is positive. $A_N$ is the number of negative actions and $Total_a$ is the total number of performed actions. $W_a$ is the action's weight according to its nature (positive, wrong, and malicious) depending upon the requesting cloud service user and performance of task ($0 \leq w_a \leq 1$).

Parameter *m* is the security level, where $m \geq 1$. This security level affects the action weight, for this reason we raise the action weight to the power of (*m*). The exponential really influences when the actions are wrong. We will show later in following diagrams how the security level affects the action values. When a new action is performed, $V_a$ is recalculated, reflecting the present behavior of the entity. The new trust value will take it into account and modify the current trust value of the service requesting CSU. If it is assumed that,

1. Initially trust value $V_a = 1$
2. $w_a$ for positive action = 1 and for malicious action = 0.8, and
3. Security level m = 1

The trust degree is updated according to action, as follows in Table I:

Table I: Representative computation of trust value for any CSU

| Iteration | Action Behavior | $A_N$ | $Total_a$ | $V_a$ |
|---|---|---|---|---|
| 1 | Positive | 0 | 1 | 1 |
| 2 | Malicious | 1 | 2 | 0.4 |
| 3 | Positive | 1 | 3 | 0.7 |
| 4 | Malicious | 2 | 4 | 0.4 |
| 5 | Malicious | 3 | 5 | 0.3 |

A CSU can perform positive or negative activity in its VM. Depending on its activity trust value in Domain Trust Table (DTT) as well as in User Trust Table (UTT) are updated. In this simulation we have chosen three types of CSUs: Trusted, Non-Trusted and Innocent. Each type of user has different probability to perform positive activity as,

- Trusted User have the probability 0.8
- Non Trusted User have the probability 0.2

- Innocent User have the probability 0.5

In this simulation the following parameters are also assumed for the two layers of the framework.

For Domain Layer (Level 1 or Broker Domain)

1. For positive activity $W_a$ =1 & negative activity $W_a$ =0.9
2. Security level m=1
3. Threshold value is 0.1

For User Layer (Level 2, Cloud Service Provider Domain)

1. For positive activity Wa =0.9 & negative activity Wa =0.8
2. Security level m=1
3. Threshold value is 0.2

We simulate the experiment for 3 types of CSU for a specific task request. We randomly choose two users for five tasks. They sent 150 requests for a particular task id 1001.

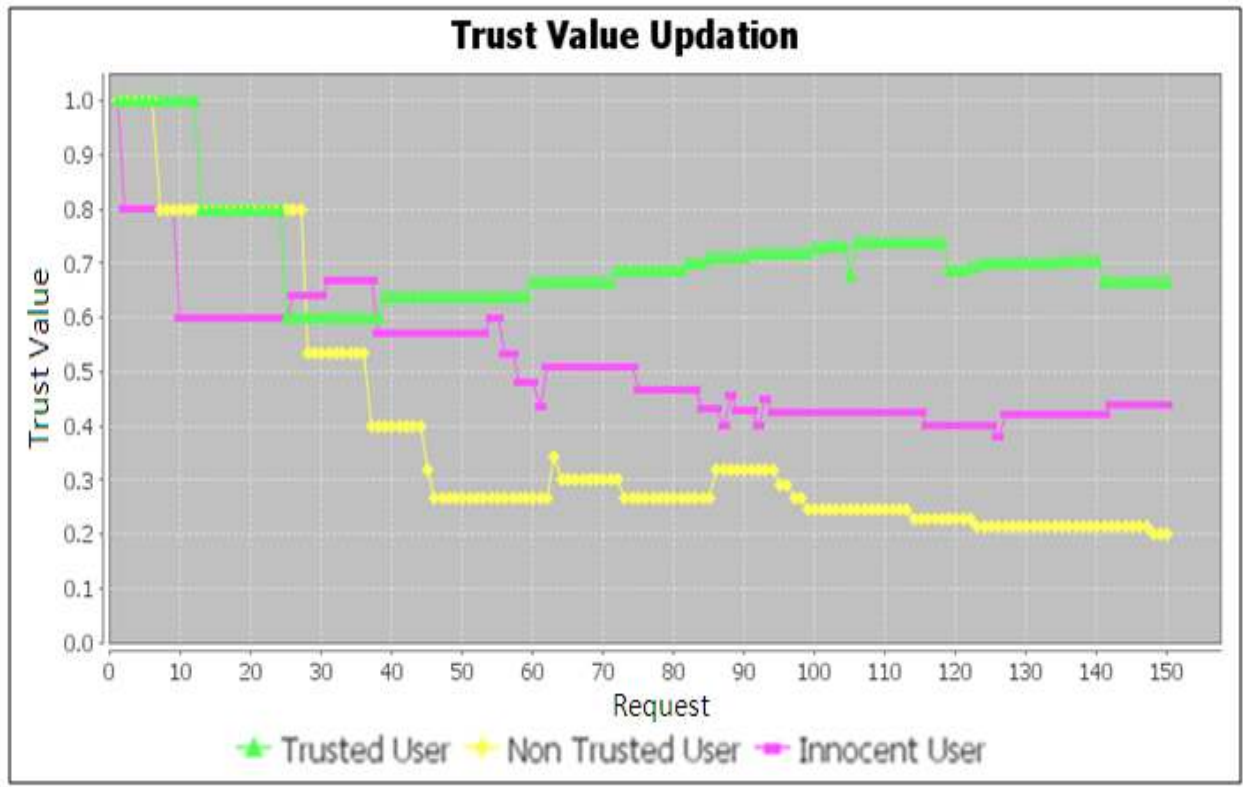

**Figure 4: Trust Value Updating - Case 1**

The above experimental result shows that the probability of doing malicious communication is much less in case of a trusted user than that of innocent and non-trusted users. We simulated several experiments in this test-simulated environment to get the results. It is clear from each simulation that the probability of reaching the threshold value in case of a non-trusted user is much higher than that of any other users. From Figure 5 it is clear that trust degree for non-trusted users is increased after performing some trusted tasks and decreases after performing malicious activities.

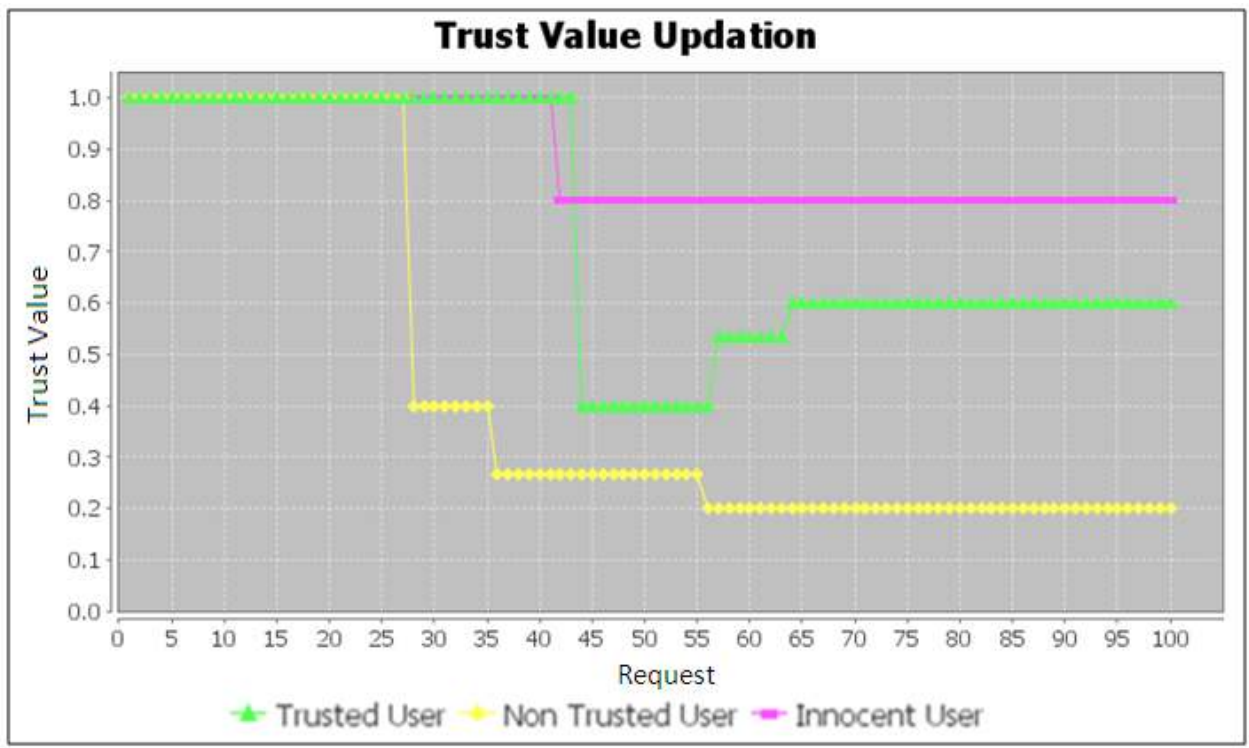

**Figure 5: Trust Value Updation- Case 2**

From Figure 5, the trust degree of an innocent cloud service user is much greater after successfully performing some trusted communication with CSP. The trust degree of trusted user decreased after some time and gradually reaches high after performing several trusted communication with the cloud service provider. The non-trusted user reaches to the threshold rapidly. In this simulation we took 3 CSU ser and 10 numbers of tasks. CSUs requested for 100 tasks with the task id 1001.

## 5. Conclusion

The proposed framework tries to maintain the domain reputation as long as possible by discarding malicious users from the domain reducing the CSP's workload. It also increases some workload of domains and this framework fails to prevent malicious activity without CSP's information. In future malicious activity identifying approach can be imposed into the proposed framework which in turn makes the system to work independent of CSP's information about malicious activity. This would help to prevent unauthorized accesses to cloud data. Research is currently going on to evaluate the performance of this framework in a real-time environment. The framework may also be extended to eradicate data leakages in a heterogeneous cloud computing platform.